\documentclass[useAMS,usenatbib]{mn2e}
\usepackage{graphicx}
\title[The Z~And-type SMC symbiotic star LIN~9]{OGLE-SMC-LPV-00861 (LIN~9): the first proven Z~And outburst in a Magellanic symbiotic star\thanks{Based on observations made with the Southern African Large Telescope (SALT) under programme 2013-1-POL\_RSA-001.}}
\author[Miszalski, Miko{\l}ajewska \& Udalski]{Brent Miszalski,$^{1,2}$\thanks{E-mail: brent@saao.ac.za} Joanna Miko{\l}ajewska$^{3}$ and Andrzej Udalski$^{4}$ \\
$^{1}$South African Astronomical Observatory, PO Box 9, Observatory, 7935, South Africa\\
$^{2}$Southern African Large Telescope Foundation, PO Box 9, Observatory, 7935, South Africa\\
$^{3}$Nicolaus Copernicus Astronomical Centre, Bartycka 18, 00716 Warsaw, Poland\\
$^{4}$Warsaw University Observatory, Al. Ujazdowskie 4, PL-00-478, Warsaw, Poland\\
}
\begin{document}

\date{Accepted . Received ; in original form }

\maketitle
\begin{abstract}
   We report on the discovery of a new Small Magellanic Cloud (SMC) symbiotic star, OGLE-SMC-LPV-00861, previously catalogued as H$\alpha$ emission line source LIN~9. The OGLE light curve shows multiple-maxima outburst behaviour over $\sim$1200 d with a maximum outburst of $\Delta V=1.5$ mag. An optical spectrum of LIN~9 taken with the Southern African Large Telescope (SALT) at quiescence reveals a K5 red giant with emission lines confirming its symbiotic star nature, demonstrating the potential use of ongoing large time-domain surveys to identify strong symbiotic star candidates. It is the first Magellanic symbiotic star proven to show poorly understood Z~And outbursts. At outburst the estimated hot component luminosity is $L\sim3165$ $L_\odot$, compared to $L\sim225$ $L_\odot$ at quiescence. Further observations are needed, especially at outburst, to better understand this unique Z And-like system at a known distance, and to provide essential input to physical models of the Z~And phenomenon.
\end{abstract}

\begin{keywords}
   surveys - binaries: symbiotic - Magellanic Clouds - stars: emission-line, Be - stars: activity - stars: individual: LIN~9
\end{keywords}

\section{Introduction}
Symbiotic stars are amongst the most variable sources in large photometric surveys (e.g. Miko{\l}ajewska 2001) owing to the rich variety of phenomena forged by the accretion of a red giant's high mass loss wind onto a white dwarf (see Miko{\l}ajewska 2012 for a recent review). Extragalactic populations of symbiotic stars offer the opportunity to study these accretion driven processes at a known distance and therefore with known luminosities of the stellar components. While there are several ongoing and successful surveys for new Galactic symbiotic stars (e.g. Corradi et al. 2008, 2010; Corradi 2012; Miszalski, Miko{\l}ajewska \& Udalski 2013; Baella, Pereira \& Miranda 2013; Miszalski \& Miko{\l}ajewska 2014; Rodr\'iguez-Flores et al. 2014), there remains a longstanding scarcity of symbiotic stars in the Magellanic Clouds. In the Small Magellanic Cloud (SMC) only seven symbiotic stars are known, LIN~358, N~60 and N~73 were found by Walker (1983), Morgan (1992) discovered SMC~1, SMC~2 and SMC~3, and the only one with warm dust surrounding a probable Mira (a dusty or D-type symbiotic, see e.g. Whitelock 1987) was found by Oliveira et al. (2013). Similarly, in the LMC only eight are known (Belczy\'nski et al. 2000). 

We have begun to investigate the use of large photometric databases to identify promising symbiotic stars, based on the presence of long periodic or outburst behaviour that can be typical of symbiotic stars. Several recent studies have used large photometric surveys such as MACHO (Alcock et al. 1997) and OGLE (e.g. Udalski 2009) to study known symbiotic stars, as well as to assist in the identification of new symbiotic stars (Miszalski et al. 2009; Gromadzki et al. 2009; Lutz et al. 2010; Kato, Hachisu \& Miko{\l}ajewska 2013; Miszalski, Miko{\l}ajewska \& Udalski 2013; Gromadzki, Miko{\l}ajewska \& Soszy\'nski 2013; Miszalski \& Miko{\l}ajewska 2014; Angeloni et al. 2014; Miko{\l}ajewska, Caldwell \& Shara 2014). We have obtained spectroscopy of several promising candidates from the OGLE-III SMC long periodic variable (LPV) catalogue (Soszy\'nski et al. 2011) with the Southern African Large Telescope (SALT; Buckley, Swart \& Meiring 2006; O'Donoghue et al. 2006). 

\begin{figure}
   \begin{center}
      \includegraphics[scale=0.4,angle=270]{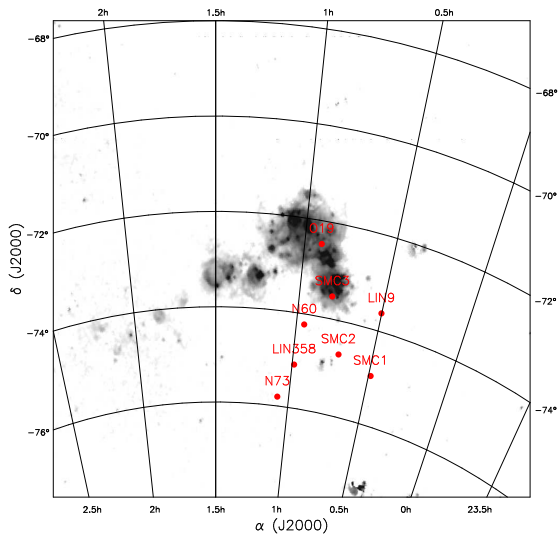}
   \end{center}
   \caption{The distribution of SMC symbiotic stars overlaid on top of the Gaustad et al. (2001) H$\alpha$ emission map in a Lambert azimuthal equal-area projection. The object O19 corresponds to the D-type symbiotic star listed as object 19 in table 1 of Oliveira et al. (2013).}
   \label{fig:dist}
\end{figure}

Here we report on the discovery of a new SMC symbiotic star, OGLE-SMC-LPV-00861 (Soszy\'nski et al. 2011), the first Magellanic symbiotic star to show Z~And outburst behaviour (e.g. Kenyon 1986). It was highlighted as an object of high interest in figure 7 of Soszy\'nski et al. (2011) and was pronounced by the authors as a rare example of an outburst in a red giant. The Two Micron All Sky Survey (2MASS, Skrutskie et al. 2006) point source catalogue designation is 2MASS J00300739$-$7337190 and it lies 19.9 arcsec from the SIMBAD coordinates of the H$\alpha$ emission line object LIN~9 (Lindsay 1961). Coordinates from old photographic plates often have large uncertainties and since there is no visible source at the SIMBAD position of LIN~9 (Hambly et al. 2001), it seems likely that LIN~9 coincides with OGLE-SMC-LPV-00861. Figure \ref{fig:dist} displays the location of LIN~9 relative to the other SMC symbiotic stars. The spatial distribution of SMC symbiotic stars highlights the challenging task remaining to identify the full population in the crowded central bar region. This has not yet been possible due to the lack of a high resolution modern H$\alpha$ survey, but this will be available in the near future (Ripepi et al. 2014).

\section{Observations}
\label{sec:obs}

\subsection{OGLE light curves}
\begin{figure}
   \begin{center}
      \includegraphics[scale=0.36,angle=270]{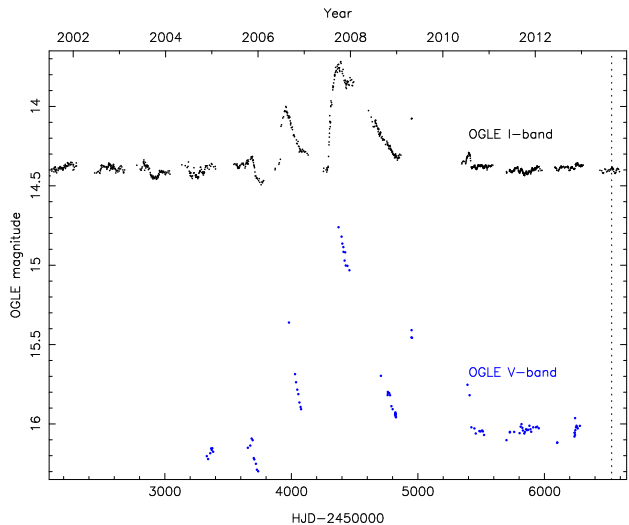}
   \end{center}
   \caption{The OGLE light curves of LIN~9. The dotted vertical line marks the epoch of the SALT RSS spectrum.}
   \label{fig:lc}
\end{figure}

Figure \ref{fig:lc} displays the OGLE-III $V$-band and $I$-band light curves (Soszy\'nski et al. 2011), updated to include final OGLE-III and recent OGLE-IV measurements. A final calibration has not yet been applied to the (post-outburst) OGLE-IV photometry (HJD-2450000 $\ge$ 5000 d), meaning the zero points of the photometry in this date range may be accurate to 0.1--0.2 mag. Table \ref{tab:phot} summarises characteristic magnitudes at quiescence and outburst we derived from the light curves. We have adopted an average $V=16.1\pm0.1$ mag at quiescence, averaging pre- and post-outburst measurements, because of the uncertain zero points. Outburst amplitudes were, however, measured directly by comparing the maximum magnitudes against the pre-outburst (HJD-245000$\sim$3740 d) average magnitudes. Similarly, quiescent $V-I$ colours were calculated at this time. The whole outburst period lasts at least $\sim$1200 days. There may be a tentative periodicity during quiescence of $\sim$500--600 days, however a better sampled $V$-band light curve is necessary to investigate this further.

\begin{table}
   \centering
   \caption{A summary of measurements made from OGLE photometry of LIN~9.}
   \label{tab:phot}
   \begin{tabular}{lrrrr}
      State & $V$ & $I$ & $V-I$ & $(V-I)_0$\\
     \hline
     Quiescence ($Q$) & 16.10: & 14.40 & 1.80 & 1.73\\
     Outburst ($O$)  & 14.75 & 13.75 & 1.00 & 0.93 \\ 
     $\Delta$ ($Q-O$)  & 1.45  & 0.65 & 0.80 & 0.80\\
     \hline
   \end{tabular}
\end{table}

\subsection{SALT RSS spectroscopy}
\label{sec:spec}

\begin{figure}
   \begin{center}
\includegraphics[scale=0.37,angle=270]{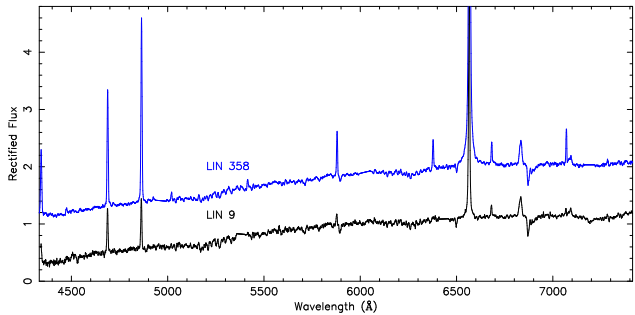}\\
   \end{center}
   \caption{SALT RSS spectra of the new SMC symbiotic star LIN~9 (bottom) and the known SMC symbiotic star LIN~358 (top). The spectra are separated by an additive constant.}
   \label{fig:smc}
\end{figure}

We observed LIN~9 with the Robert Stobie Spectrograph (RSS; Burgh et al. 2003; Kobulnicky et al. 2003) on the queue-scheduled SALT under programme 2013-1-POL\_RSA-001 (PI: Miszalski). An exposure of 1755 s was taken on 2013 August 27 with an identical RSS configuration to that described in Miszalski \& Miko{\l}ajewska (2014). After basic reductions were applied using the \textsc{pysalt} package (Crawford et al. 2010), the same data reduction process was also employed, resulting in the spectrum displayed in Figure \ref{fig:smc}. 

Included in Fig. \ref{fig:smc} is a SALT RSS spectrum of another SMC symbiotic star, LIN~358 (Lindsay 1961; Walker 1983). In this case the PG900 grating was used on 2013 September 23 with a 1.5 arcsec wide slit to take exposures of 1850 s ($\lambda=$3915--7010 \AA) and 1500 s ($\lambda=$6164--9160 \AA). Contemporaneous spectroscopic flat-fields were taken with each spectrum and were applied to the science exposures after each flat was divided by a 100 pixel median of itself to reduce the amplitude of detector fringing patterns and to correct bad pixels. Cosmic ray events were cleaned using the \textsc{lacosmic} package (van Dokkum 2001). Wavelength calibration, one-dimensional spectral extraction and flux calibration was performed as in Miszalski \& Miko{\l}ajewska (2014). The overlap region between the two spectra was used to scale and then combine the two spectra together. Both spectra are normalised near $\sim$6300\AA\, since the moving pupil design of SALT does not allow for an accurate absolute flux scale.

\section{Results and Discussion}
\subsection{Spectral features and reddening}
\label{sec:featred}
The spectrum of LIN~9 was taken during quiescence (Fig. \ref{fig:lc}) and satisfies the criteria for a symbiotic star (Belczy\'nski et al. 2000), namely that the cool giant is present, the hot component is confirmed with the presence of He~II $\lambda$4686 emission, and the telltale Raman-scattered O~VI emission bands (Schmid 1989). We have applied an absolute flux scale to the SALT RSS spectrum using the adopted average $V=16.1$ mag such that convolution of the spectrum with the Johnson $V$ filter agrees with the $V$ mag. Table \ref{tab:lines} gives the emission line fluxes that were measured by fitting Gaussian profiles to the calibrated spectrum. 

\begin{table}
   \centering
   \caption{Emission line fluxes in $10^{-15}$ erg cm$^{-2}$ s$^{-1}$ measured from the SALT spectrum scaled to $V=16.1$ mag.}
   \label{tab:lines}
   \begin{tabular}{lr}
      Line & Flux\\
      \hline
     {} He~II $\lambda$4686 & 8.3\\
     {} H$\beta$ 4861 & 9.8\\
     {} He~I 5876 & 2.2\\
     {} H$\alpha$ 6563 & 54.6\\
     {} He~I 6678 & 1.9\\
     {} O~VI 6830 & 8.0\\
     {} He~I 7065 & 1.3\\
     {} O~VI 7088 & 2.1\\
      \hline
   \end{tabular}
\end{table}

Since Case B conditions do not apply for S-type symbiotic stars, the Balmer decrement cannot be used to estimate the reddening. The reddening can be assumed to be less than the average for known SMC symbiotic stars ($A_V\la0.3$ mag; M\"urset et al. 1996) and greater than the total galactic contribution $A_V\sim0.04$ mag (Schlafly et al. 2011). This range can be refined further by taking the total HI column density of the SMC at the position of LIN~9, namely $N(HI)$=5--10$\times$10$^{20}$ cm$^{-2}$ (Luks 1994), corresponding to $A_V=N(HI)/(7.4\times10^{21})\sim$0.07--0.14 (Gordon et al. 2009). Assuming LIN~9 is located behind the HI cloud, we then obtain $A_V\la 0.04+0.14=0.24$ mag. Furthermore, the 2MASS $(J-K_s)$ colour of LIN~9 is almost identical to LIN~358 which has an (uncertain) estimate of $E(B-V)\sim0.08$ or $A_V\sim0.24$ mag (M\"urset et al. 1996). We therefore adopt $A_V\sim$0.10--0.24 mag, i.e. $A_V=0.17\pm0.07$ mag or $E(V-I)=0.07\pm0.03$ mag for LIN~9. 

\subsection{Spectral classification of the cool component}
The cool component of LIN~9 bears close resemblance to the K-type giant of LIN~358 (M\"urset et al. 1996; Fig. \ref{fig:smc}). In quiescence the $(V-I)_0\sim1.7$ mag colour of LIN~9 corresponds to a K5 giant (Bessell \& Brett 1988). The luminosity of the red giant can be estimated from its NIR 2MASS magnitudes ($J=13.29$, $H=12.44$ and $K_s=12.27$ mag). The bolometric correction is estimated to be $BC(K)=2.61$ for $J-K=1.02$ (Bessell \& Wood 1984). When combined with $K_0=$12.25 mag, we find $M_{bol}$=$-$4.0 mag and $L=3311$ $L_\odot$. The effective temperature of the giant can be estimated using the colour-T$_\mathrm{eff}$ relations for metal-poor giants calculated by Ku{\v c}inskas et al. (2006). Assuming [Fe/H]=$-1$, the $J-K$ colour of LIN~9 is then consistent with T$_\mathrm{eff}$=3900 K. We then estimate the red giant radius $R=126$ $R_\odot$ assuming a distance of $60.6\pm1.0$ kpc (Hilditch et al. 2005). This is similar to those of other SMC symbiotic star giants (M\"urset et al. 1996).

\subsection{The nature of the hot component}
\label{sec:hot}
During outburst we can estimate the magnitudes and $V-I$ colour of the hot component by subtracting the quiescent (pre-outburst) $V$ and $I$ magnitudes that represent the cool companion (16.2 and 14.4 mag, respectively). We obtain $V=15.15$, $I=14.60$, and $V-I=0.55$ mag. The dereddened $V-I=0.48\pm0.03$ mag is consistent with an A or F-type giant or supergiant (Cousins 1980), assuming that the optical bands are dominated by the A or F-type photosphere, rather than a nebular continuum which should much redder, $(V-I)_\mathrm{neb}\sim0.8-0.9$.\footnote{We used an estimate of filter average magnitudes for free-free and bound-free hydrogen emission, assuming case B recombination and $T=$10--20 kK. If we assume the continuum during the optical maximum were dominated by nebular emission, then $V\sim15$ mag would require an enormous emission measure (EM) of a few times $10^{61}$ cm$^{-3}$, 1--2 orders of magnitude larger than nebular emission in any of the Z And-type outbursts. For example, Skopal et al. (2009) estimated EM=$6\times10^{59}$ cm$^{-3}$ during the hot outburst of AG Dra when the EM was very high. Moreover, to produce such an emision measure the hot component luminosity would have to exceed the Eddington limit of a 1.0 $M_\odot$ white dwarf.} So, assuming that most of the hot component continuum emission is shifted to the optical (a lower limit to $L$ if not), a distance of $60.6\pm1.0$ kpc (Hilditch et al. 2005), and that during outburst $m_\mathrm{bol}\sim V_\mathrm{hot}$ (i.e. BC~0 for an A or F-type giant or supergiant, see e.g. Straizys \& Kuriliene 1981), we estimate $M_\mathrm{bol}\sim-4.0\pm0.1$ mag which corresponds to $\sim3165$ $L_\odot$.

At quiescence we make use of the fluxes in Tab. \ref{tab:lines} to estimate the temperature and luminosity of the hot component. The minimum temperature is set by the maximum ionization potential (IP) observed in the spectrum, namely IP=114 eV, and the relation $T/10^3\sim \mathrm{IP (eV)}$ found by M\"urset \& Nussbaumer (1994), to give $T\ga114$ kK. An upper limit for $T$ of 170 kK was derived from H$\beta$, He~II $\lambda$4686 and He~I $\lambda$5876 emission line ratios assuming Case B recombination (Iijima 1981). We estimate the luminosity of the hot component to be $L\sim225$ L$_\odot$ using equation 8 of Kenyon et al. (1991). Similarly, equations 6 and 7 of Miko{\l}ajewska et al. (1997) give $L(\mathrm{He II}\ \lambda4686)\sim366$ $L_\odot$ and $L(H\beta)\sim219$ $L_\odot$. Several other symbiotic stars in Miko{\l}ajewska et al. (1997) were found to have similar luminosities for their hot components. All these estimates assume blackbody spectrum for the hot component, $d=60.6$ kpc (Hilditch et al. 2005), case B recombination for the emission lines, and are accurate to only a factor of $\sim$2. 

\subsection{Z~And outburst classification}
The outburst amplitudes observed in LIN~9 (Tab. \ref{tab:phot}) are comparable to typical amplitudes of the optical outbursts in classical Z~And-type symbiotic stars (e.g figure 1.2 of Kenyon 1986; see also figure 4 of Kenyon et al. 1991). The outburst behaviour observed in symbiotic stars, based on their activity, can either occur in ordinary or classical symbiotic stars (e.g. Z~And, CI~Cyg, AX~Per, and AG~Dra as typical examples), or in symbiotic novae that are essentially thermonuclear novae. Many of these classical systems, including Z~And (the prototype), CI~Cyg, and AX~Per, show occasional 1--3 mag optical eruptions with timescales from months to years. During this time the hot component maintains roughly constant luminosity (within a factor of a few), whereas its effective temperature decreases to $\sim$10$^4$ K. The mechanism of this activity has been only recently explained by unstable disc-accretion onto hydrogen-shell burning white dwarf (Miko{\l}ajewska 2003; Sokoloski et al. 2006). There still remains, however, no quantitative model of the phenomenon. 

Another argument in favour of the Z~And-type outburst is the lack of a rich nebular spectrum after the optical maximum, especially one containing forbidden lines. Such a spectrum is present in all known SyNe once they decline from their optical maximum. Although Galactic Z~And-type systems also show some forbidden lines (e.g. [O~III], [Ne~III], [Fe~VII]), these lines are never as strong as in SyNe, and they crucially do not show the lower density lines (e.g. [O~II], [S~II], [N~II]; see e.g. CI~Cyg - Kenyon et al. 1991; AG~Dra - Miko{\l}ajewska et al. 1995; Z~And - Miko{\l}ajewska \& Kenyon 1996). This should also hold for LIN~9 in the low metallicity environment of the SMC, since similarly rich nebular spectra are also observed in classical novae in the SMC (e.g. Nova SMC 2001, Mason et al. 2005).

\subsection{Metallicity effects}
The LIN~9 spectrum is very similar to the quiescent optical spectrum of AG~Dra (Miko{\l}ajewska et al. 1995; Munari et al. 2009), the Galactic halo Z~And-type symbiotic with a very low metallicity of [Fe/H]=$-$1.3 dex (Smith et al. 1996). In particular, in both LIN~9 and AG~Dra, there are no forbidden emission lines in the optical spectra, except for relatively faint auroral [OIII] $\lambda$4363 and [Ne~III] $\lambda$3869 lines present in some spectra of AG~Dra (Miko{\l}ajewska et al. 1995). Significantly, there are no high ionization [Fe~VII] $\lambda\lambda$6089, 5721 emission lines that usually accompany the Raman scattered O~VI lines (Schmid 1989) in spectra of Galactic symbiotic stars (Allen 1984; Miko{\l}ajewska, Acker \& Stenholm 1997; Munari \& Zwitter 2002; Miszalski \& Miko{\l}ajewska 2014). The lack of [Fe~VII] and other metallic lines in AG~Dra, as well as some other SMC symbiotic systems (e.g. Allen 1984; Munari \& Zwitter 2002), may be symptomatic of their low metallicity. This may also be the case for the carbon-rich Galactic symbiotic 2MASS J16003761$-$4835228 which displays coronal [Fe~X] $\lambda$6735 emission, but not [Fe~VII] emission (Miszalski \& Miko{\l}ajewska 2014).

\section{Conclusions}
\label{sec:conclusion}
We have discovered a new SMC symbiotic star, LIN~9, with SALT RSS spectroscopy of a remarkable candidate first highlighted by Soszy{\'n}ski et al. (2011) for its unusual outbursts in the OGLE light curve. We have proven that the outburst event cannot be ascribed to a symbiotic nova, but instead must be a Z~And outburst. Further study of this unique Z~And Magellanic symbiotic star at a known distance will help in the development of physical models for the still poorly understood Z~And outburst mechanism. The discovery of LIN~9 brings the total number of symbiotic stars known in the SMC to eight and demonstrates that spectroscopic follow-up of candidates that exhibit long term periodic behaviour or outbursts can indeed help alleviate the current paucity of identified Magellanic symbiotic stars in the near future.

\section{Acknowledgements}
Some of the observations reported in this paper were obtained with the Southern African Large Telescope (SALT) and we would like to thank the Polish and South African SALT time allocation committees for the award of SALT time. We thank the anonymous referees for constructive reports including suggestions that helped improve this work. JM is supported by the Polish National Science Center grant number DEC-2011/01/B/ST9/06145. The OGLE project has received funding from the European Research Council under the European Community's Seventh Framework Programme (FP7/2007-2013) / ERC grant agreement no. 246678 to AU. This research has made use of the SIMBAD database and VizieR catalogue access tool, operated at CDS, Strasbourg, France. This research has made use of SAOImage DS9, developed by Smithsonian Astrophysical Observatory. This publication makes use of data products from the Two Micron All Sky Survey, which is a joint project of the University of Massachusetts and the Infrared Processing and Analysis Center/California Institute of Technology, funded by the National Aeronautics and Space Administration and the National Science Foundation.

\end{document}